\documentclass[preprint,12pt]{article}
\usepackage[utf8]{inputenc}

\usepackage{amsmath}
\usepackage{amsfonts}
\usepackage{amssymb}
\usepackage{graphicx}
\usepackage{hyperref}
\usepackage{authblk}

\begin{document}

\title{Wavelength selection beyond Turing}

\author[1]{Yuval R. Zelnik}
\author[2]{Omer Tzuk}
\affil[1]{Centre for Biodiversity Theory and Modelling, Station d'Ecologie Experimentale du CNRS, Moulis 09200, France}
\affil[2]{Physics Department, Ben-Gurion University of the Negev, Beer Sheva 8410501, Israel}

\maketitle
\begin{abstract}
       Spatial patterns arising spontaneously due to internal processes are ubiquitous in nature, varying from regular patterns of dryland vegetation to complex structures of bacterial colonies. Many of these patterns can be explained in the context of a Turing instability, where patterns emerge due to two locally interacting components that diffuse with different speeds in the medium. Turing patterns are multistable, such that many different patterns with different wavelengths are possible for the same set of parameters, but in a given region typically only one such wavelength is dominant. In the Turing instability region, random initial conditions will mostly lead to a wavelength that is similar to that of the leading eigenvector that arises from the linear stability analysis, but when venturing beyond, little is known about the pattern that will emerge. Using dryland vegetation as a case study, we use different models of drylands ecosystems to study the wavelength pattern that is selected in various scenarios beyond the Turing instability region, focusing the phenomena of localized states and repeated local disturbances.
\end{abstract}

\section{Introduction}
\label{intro}
Spatial patterns arise in various natural systems, including clouds, bacterial colonies, animal coats, and dryland ecosystems \cite{Rietkerk2008tree}. Many of these patterns are periodic, exhibiting a typical length-scale. However, in most of these systems, and the theoretical models suggested to describe them, more than one wavelength is possible for the same set of parameters. The periodic patterns observed in many of these systems can be explained as a result of a Turing instability \cite{Turing1990chemical,CrossHohenberg,Borckmans1995turing}, where the dynamics of an activator and inhibitor that react locally and diffuse in space in different rates leads to the rise of spatial patterns. These so called Turing patterns show a spectrum of possible wavelengths of stable patterns, and therefore the observed wavelength typically depends on the history of the system. 

A question the often arises is what causes the system to converge to a specific wavelength. This question of wavelength selection has often focused on the "Turing regime", where a uniform state is unstable to small non-uniform perturbations (but stable to uniform ones). In this context, one can make a linear approximation around the uniform state and infer what patterns will arise from small perturbations around that uniform state \cite{CrossHohenberg,Borckmans1995turing}. In these conditions one may expect the patterns to follow the wavelength that corresponds to the fastest growing mode, although this may not be the case due to the non-linearities of the system \cite{Schober1986dynamics}. However, large perturbations may have markedly different effects than the small ones that conform to linear analysis \cite{Dee1983propagating}. Moreover, periodic patterns are often bistable with a uniform state in some parameter range, where such predictions are not relevant since the uniform state in question is stable by definition \cite{Metens1997pattern,Judd2000simple}. 

One interesting aspect of wavelength selection, arising in systems where the periodic patterns are bistable with a uniform state, is that of localized states \cite{Coullet2000stable,Burke2006localized}. These states are a spatial mix of a periodic state and a uniform one, that are stable in some parameter range. Recent interest has sparked on different aspects of these localized states, and their bifurcation structure, often termed homoclinic snaking \cite{Burke2007homoclinic}. The quasi-periodic domain of the localized states shows a specific wavelength, and as such it brings about the question of how this wavelength is selected \cite{Burke2007homoclinic,Knobloch2008spatially}. In particular, the relation between this snaking wavelength and system parameters is largely unclear in systems with no Hamiltonian functional \cite{Knobloch2008spatially} of the stationary problem, which includes many realistic systems where patterns occur.

One field where periodic patterns are particularly interesting is dryland ecosystems \cite{Valentin1999soil,Klausmeier1999regular}. Periodic patterns have been observed in drylands throughout the world \cite{Valentin1999soil,Deblauwe2008global}, and many models have been proposed to describe these patterns, using several distinct physical mechanisms \cite{Lefever1997bmb,Klausmeier1999regular,Hillerislambers2001vegetation,Gilad2004prl}. The emergence of pattens in dryland ecosystem is associated with mechanisms of positive feedback between local vegetation growth and water transport towards the location of the vegetated patch \cite{Meron2015book}. Along with feedbacks that account for the creation of patterns, dryland ecosystem often also have certain ranges of bistability  or multistability of the possible states of the system: bare-soil, uniform-vegetation, and non-uniform patterned states. 

The question of pattern wavelength has been of high interest in this field \cite{Deblauwe2008global,Penny2013local,Sherratt2015using,Siero2015striped} since it may allow for assessment of ecosystem properties from remote sensing observations, give indicators for regime shifts and desertification, that can be relevant for efforts to intervene in degraded landscapes. Several studies that have looked at wavelength selection in dryland ecosystems \cite{Sherratt2013history,Dagbovie2014pattern,Siteur2014beyond,Siero2015striped}, have focused on how the pattern changes with a slow change in a bifurcation parameter, typically precipitation. In this context the Busse balloon \cite{Busse1978non}, the parameter space of wavelength and a bifurcation parameter showing the existence and stability of periodic solutions, is often helpful in predicting the dynamics in such systems. Indeed, it has been used to show how a given history of parameter change, with and without noise, can affect the pattern that is selected \cite{Siteur2014beyond,Siero2015striped}. Localized states and homoclinic snaking have been recently found in several models of dryland ecosystems \cite{Zelnik2015pnas,Dawes2015localised,Zelnik2016localized}, but the question of which wavelength the localized state take has been largely overlooked \cite{Zelnik2016desertification}.

In this paper we will focus on wavelength selection beyond the Turing regime, using models of dryland ecosystems as a case study. We will look beyond the Turing regime in two manners, by moving beyond the Turing instability region to bistable regions, and by looking at the effect of disturbances that are not locally small, and therefore can not be linearly approximated. We limit the study to one dimensional systems, as this eases the numerical analysis and keeps us away from related but different issues such as pattern directionality and frustration. To show the generality of results, we will use analysis and simulations from three different models of dryland vegetation. In the first subsection of the results section we will discuss the general structure of a typical bifurcation diagram for these models, and its repercussions. We will then look at the wavelength of localized states in a bistability range in the second subsection. Finally in the third subsection of the results we will look at the patterns observed inside the Turing range and outside of it, focusing on the effect of repeated local disturbances. 

\section{Methods}
\label{sec:methods}
To test the properties of wavelength selection we have looked at a range of five different models of dryland vegetation where patterns arise due to dynamics between vegetation and water. We choose to describe and present results from three of these models, that we believe demonstrate well the general behaviour of dryland ecosystems with vegetation patterns. These models track the dynamics of the vegetation biomass $b$, and the underground water $w$, with one model also tracking the overland water $h$. The mechanisms that are included in the models differ, and hence also their level of complexity. We limit the present work to one-dimensional systems, as its analysis and numerical investigation is more accommodating, and we believe the general trends we describe should hold for two-dimensional systems as well.

\subsection{The Rietkerk (R) model}
\label{sec:Rmodel}
The non-dimensional equations of the model presented by Rietkerk et al. \cite{rietkerk2002self} is:
\begin{subequations}
\begin{align}
    \partial_t b &= \frac{w}{w+1} b - \mu b + \nabla^2 b\\
    \partial_t w &= \alpha \frac{b+f}{b+1} h - \nu w - \gamma \frac{w}{w+1} b + D_w \nabla^2 w \\
    \partial_t h &= p - \alpha \frac{b+f}{b+1} h + D_h \nabla^2 h \, .
\end{align}
\end{subequations}
The term $\frac{w}{w+1}$ is the growth rate of the biomass, that has functional form of Holling type II with respect to the water variable, $\mu$ is the biomass decay rate, and seed dispersal or clonal growth is captured by $\nabla^2 b$. The R model brings into account the increased infiltration of surface water into the soil in patches of vegetation growth, that creates a positive feedback between vegetation growth and underground water availability. This feedback is captured by the term $\alpha \frac{b+f}{b+1}$, where the parameter $f$ stands for the infiltration contrast between bare-soil and vegetated patches, and $\alpha$ is the maximum infiltration rate into the soil in fully vegetated patch. The parameters $\nu$ and $\gamma$ are the evaporation and transpiration coefficients respectively, and the underground water diffusion parameter is $D_w$. In the overland water equation, $p$ is the precipitation parameter and $D_h$ the overland water diffusion coefficient. The branch of uniform vegetated state bifurcates supercritically for any choice of the model parameters, therefore there is no bistability of the two uniform states in this model. 

The results shown for the R model were done using the parameters given in Table \ref{tab:R_Ps}.

\begin{table}
\caption{Parameter sets for the R model.}
\label{tab:R_Ps}       
\begin{tabular}{llllllll}
\hline\noalign{\smallskip}
set & $\mu$ & $\nu$  & $\gamma$ & $\alpha$ & $f$   & $D_w$ & $D_h$  \\
\noalign{\smallskip}\hline\noalign{\smallskip}
1   & 0.2 & 1.0  & 0.05   & 0.5    & 0.2 & 1.0 & 1000 \\
2   & 0.2 & 1.0  & 0.05   & 0.5    & 0.2 & 1.0 & 2300 \\
3   & 0.9 & 0.25 & 0.35   & 1.5    & 0.2 & 100 & 3000 \\ 
\noalign{\smallskip}\hline
\end{tabular}
\end{table}
\subsection{The simplified Gilad (G) model}
\label{sec:Gmodel}
We use a simplified version of a model by Gilad et al. \cite{Gilad2004prl}, where we assume that both the roots are confined in the lateral direction and that that infiltration of water into the soil is fast and homogeneous across space \cite{Zelnik2016localized}. The combination of fast soil-water diffusion and strong water update by the plants creates the positive feedback required for destabilization of the uniform vegetated states. With these assumptions we can omit the dynamics of overland water flow, which leaves us with a two variables system, that in its non-dimensional form read
\begin{subequations}
\begin{align}
    \partial_t b &= \lambda w b (1+\eta b)^2 (1-b) -b + \nabla^2 b \\
    \partial_t w &= p - \nu w (1 - \rho b) - \lambda w b (1+\eta b)^2 + \delta_w \nabla^2 w \, ,
\end{align}
\end{subequations}
where $\lambda$ is the growth rate parameter, and $\eta$ is related to the root to shoot ratio. In the water equation, $p$ is the precipitation, $\nu$ is the evaporation term, $\rho$ captures the shading feedback that reduces the evaporation rate, and $\delta_w$ is the scaled water diffusion coefficient. Despite the fact that the G model neglects the overland water dynamics, it has a higher flexibility over the control of positive feedback mechanisms for the uniform states, and therefore we can find either a subcritical or supercritical bifurcation of the uniform-vegetation state from the bare-soil state.

The results shown for the G model were done using the parameters given in Table \ref{tab:G_Ps}.
\begin{table}
\caption{Parameter sets for the G model.}
\label{tab:G_Ps}       
\begin{tabular}{llllll}
\hline\noalign{\smallskip}
set & $\lambda$ & $\eta$ & $\nu$ & $\rho$ & $\delta_w$  \\
\noalign{\smallskip}\hline\noalign{\smallskip}
1   & 2.0     & 6.0  & 2.0 & 0.2  & 1000     \\
2   & 1.0     & 6.0  & 2.0 & 0.2  & 1000     \\
2   & 0.46  & 2.52  & 1.43 & 0.7  & 125     \\
3   & 1.0     & 1.5  & 0.5 & 0.0  & 1750     \\
\noalign{\smallskip}\hline
\end{tabular}
\end{table}

\subsection{The modified Klausmeier (K) model}
\label{sec:Kmodel}
We use a modified version of the model introduced by Klausmeier \cite{Klausmeier1999regular}. This model adds a water diffusion term instead of a water advection one with respect to the original model \cite{van2013rise}, and its equations, expressed in their non-dimensional form, are
\begin{subequations}
\begin{align}
    \partial_t b &= w b^2 - \mu b + \nabla^2 b \\
    \partial_t w &= p - w  - w b^2 + \delta_w \nabla^2 w \,.
\end{align}
\end{subequations}
The model captures the local facilitation effect of vegetation growth by using a non-linear dependency of the growth rate by the biomass, thus creating a pattern-forming feedback similar to the G model. The three parameters of this model $p$,$\mu$, and $\delta_w$, are the precipitation, biomass decay rate, and water diffusion coefficient respectively. This model has the property where the bare-soil state is always stable, so that we explore in it the context of tristability, of bare-soil, uniform-vegetation, and periodic patterns.

We used the following set of parameters for this model:
\begin{gather}\label{park1}
\text{ $\mu = 0.5$, $\delta_w = 7.5$.}
\end{gather}

\subsection{Analysis of steady-state solutions}
\label{sec:steady_state}
The first length scale that we take into account is the dominant wavelength of the system $\lambda_d$. A Turing instability occurs when a uniform stable state becomes unstable to non-uniform perturbation with a non-zero wavenumber. Linear stability analysis gives us the dependency of the growth rate of small perturbations on their wavelength. A system undergoes a Turing instability when the growth rate of small perturbation become non-zero at a finite wavelength. Beyond the Turing instability point the system may be either linearly stable to non-uniform perturbations, or become unstable to a range of wavelengths that include $\lambda_d$. Nevertheless, linear stability analysis still yields the wavelength of the perturbation with the fastest growth rate (even if it is negative).

A Second length scale arises from the eigenvalue problem of the spatial dynamics representation of the model. Looking at the steady state of the system $ 0 = \partial_t \mathbf{u} = \mathcal{R}(\mathbf{u};\mathbf{p}) + \mathbf{D} \partial_{xx} \mathbf{u} $, where $\mathcal{R}$ contains the local terms and $\mathbf{D}$ is the diagonal matrix of diffusion coefficients, we reduce the second order differential equation to a set of first order equations:
\begin{subequations}
\begin{align}
    \partial_x \mathbf{u} &= \mathbf{u}_x \\
    \partial_x \mathbf{u}_x &= - \mathbf{D}^{-1}  \mathcal{R}(\mathbf{u};\mathbf{p}).
\end{align}
\end{subequations}\label{spatial_problem}
Linearization of this system gives the spatial eigenvalues of the problem $\mu_j$. In the Turing regime all the eigenvalues lie on the imaginary axis and are related to the oscillatory mode of the system. Complex eigenvalues are associated with damped oscillations that connect the periodic domain to a uniform domain. We denote the length scale that is associated with the (absolute) imaginary valued spatial eigenvalue as the tail wavelength $\lambda_t = |1 / \rm Im (\mu_j)|$. 

The oscillatory domain wavelength of the localized solutions along the homoclinic snaking branches is different from the Turing wavelength. In order to estimate the wavelength of the snaking solutions we used AUTO \cite{auto} to perform numerical continuation of the different models. We detected the bifurcations to homoclinic snaking branches, and extracted the profiles of the solutions along these branches. Measuring the mean distance between the dips gave us another characteristic wavelength, the snaking wavelength $\lambda_s$, that is associated with the homoclinic snaking solutions. 

The bifurcation diagrams and profiles of solutions were found using numerical continuation with AUTO \cite{auto}, where the stability of branches was found using numerical stability analysis using extracted solutions, in a periodic domain of 10 periods of each solution, with periodic boundary conditions. The Busse balloons we found using numerical continuation with a proprietary code using Matlab, where the stability information was found using numerical stability analysis in a domain of 10 periods, with periodic boundary conditions.

\subsection{Dynamical simulations}
\label{sec:dynamical}
We consider two scenarios of dynamical simulations in this paper using the G and R models: random initial conditions and repeated local disturbances. In both scenarios we apply time integration using a pseudo-spectral method with periodic boundary conditions. At the end of each simulation we count the number of peaks/gaps and divide the system size by this number to give us the effective wavelength. This definition coincides with the normal definition of wavelength for periodic solutions, and we find it more informative for the localized solutions. The system size is chosen to accommodate large number of periods so that boundary conditions effects are negligible, and is 800 and 3200 for the G and R models respectively. All the results were averaged over 100 simulations using different randomization seeds. 

In the first scenario we start with random initial conditions and let the system run for a predefined time of $t=100$ and $t=500$ for the G and R models respectively, until the system reaches an approximate steady-state. We choose to initialize the state with biomass values from a uniform random distribution with no spatial correlations, while the water variables start with values corresponding to the bare-soil solution. The biomass values range from $0$ in both models, to either $0.7$ or $2.0$ for the G and R models respectively. The goal in mind is to have the same initial conditions for all values of precipitation (where for some of the range the bare-soil state may not be stable, and for others the uniform-vegetation is not stable or does not exist), so we might compare the results with minimal bias. Using this method, for the precipitation range where the uniform-vegetation is Turing unstable we get results that are comparable to those of small perturbations, while inside the bistability regions we get spatial patterns for much of the range. While this method depends on the arbitrary choice of maximal biomass values (which was chosen to be comparable to the typical maximal biomass in each model), we have chosen it as it allows us to make a consistent comparison between this scenario and the effect of repeated local disturbances.

In the second scenario we start with the same initialization and time integration as the previous scenario, then we repeatedly perturb the system by local disturbances with a constant relaxation time between them, and finally we let the system converge to a new steady state. In the interim stage we repeatedly disturb the system by taking out biomass from a confined domain of size $s$ (the disturbance size) in a randomly chosen location, and then wait for a relaxation time of $\tau=10$ and $\tau=50$ for the G and R models respectively. This is repeated 500 times, so that the system effectively converges to a new steady-state (that is, using more repetitions does not affect the qualitative results). In the final stage the system is run again for the same convergence time as in the first stage. Note that the choice of $\tau$ may be significant in part of the parameter range, where a very short relaxation time will push the system to a bare-soil state if it is stable. We choose $\tau$ that is large enough so that using a longer relaxation time does not affect the qualitative results.

\section{Results}
\label{sec:results}
\subsection{Bifurcation structure and multistabilty}
\label{sec:bifurcations}
Dryland vegetation models reproduce many of the spatial patterns observed in various ecosystems, with bare soil, patterned vegetation, and uniformly distributed vegetation along an increasing precipitation gradient. Depending on the types of positive feedbacks between water availability and vegetation growth, the periodic pattern may be bistable in part of its range with a bare-soil state, a uniform-vegetation state, or both. In order to demonstrate the wavelength selection in various conditions, we focus on parameters that lead to patterns which are stable with both bare-soil and uniform-vegetation states. 

\begin{figure}
    \centering
    \includegraphics[scale=0.4]{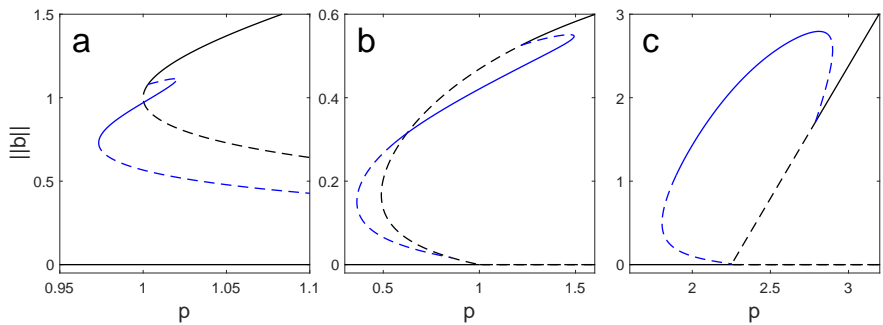}
    \caption{Three general structures of bifurcation diagrams that are examined, obtained from different models.\textbf{(a)} The K model showing a tristability of bare-soil, uniform-vegetation and patterns. \textbf{(b)} The G model with parameter set 1 showing a uniform-vegetation branch bifurcating subcritically from the bare-soil one without a tristability range. \textbf{(c)} the R model with parameter set 3 showing a supercritical bifurcation of the uniform-vegetation branch from the bare-soil one. The bare-soil and uniform states are colored in black, while one of the many periodic states is shown in blue. Solid (dashed) lines denote stable (unstable) states.}
    \label{fig:bif3}
\end{figure}

If the bifurcation point between bare-soil and uniform-vegetation is a supercritical one, then there are always two separate bistability ranges, of patterns with bare-soil and of patterns with uniform-vegetation. However, if the bifurcation is subcritical then there are either two separate bistability ranges as before, or they merge to form a tristability range of bare-soil, uniform-vegetation, and periodic patterns. In order to study the dynamics that affect wavelength selection of the solutions, we will compare three general structures of the bifurcation diagrams: (1) bifurcation diagram in which the uniform vegetates state bifurcates supercritically from the bare soil state, and subsequntly looses its stability to non-uniform branch, (2) bifurcation diagram in which there is tristability between the uniform vegetation, non-uniform vegetation, and homoclinic snaking solution, and (3) an intermediate type of bifurcation diagram in which the uniform vegetation is subcritical. We demonstrate these three possibilities in Fig. \ref{fig:bif3}, where we show three bifurcation diagrams of the three models we study. 

\begin{figure}
    \centering
    \includegraphics[scale=0.4]{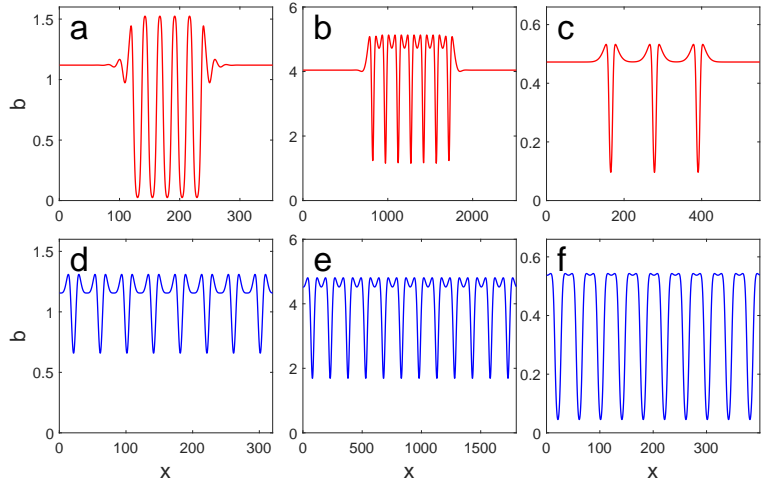}
    \caption{Solution profiles of localized state (top row) and periodic states (bottom row).  Models used were: \textbf{(a,d)} K model, \textbf{(b,e)} R model with parameter set 1, \textbf{(c,f)} G model with parameter set 4. \textbf{(a)} The localized state shows large $\lambda_s$ due to proximity to the Maxwell point. \textbf{(c)} The localized state shows large $\lambda_s$ due to the large tail wavelength (see main text for details).
    }
    \label{fig:profiles}
\end{figure}

The bistability range of patterns with uniform-vegetation often contains a subrange where localized states, a spatial mix of periodic pattern and a uniform-state, exist and are stable. On the other hand, the bistability range of patterns with bare-soil does not typically show such a range \cite{Zelnik2016desertification}. Within the snaking range, the localized states have a specific wavelength per set of parameters, unlike the periodic patterns that show multiple possible wavelengths (see Fig. \ref{fig:profiles}). The patterns observed within the snaking range are often dominated by the localized states, so that the we might assume that this snaking wavelength is significant in the pattern selection that occurs inside the bistability range of patterns and uniform-vegetation. The question of what affects the snaking wavelength and how it is chosen has been addressed in several models, and was shown to be related to an spatial Hamiltonian \cite{Burke2007homoclinic}. Dryland ecosystems are however a dissipative system where such a functional does not exist or is not known, so that the question of how the snaking wavelength is selected remains unanswered. In the following subsection we will consider three other length scales (that arise in such systems) that may affect the snaking wavelength in different circumstances. 

\subsection{Snaking wavelength}
\label{sec:snaking_wl}
We will consider here three different possible mechanisms that may affect the selection of the snaking wavelength. These can be associated with three general types of states in the system, the periodic pattern from which the localized states need to "choose" a wavelength, the uniform-vegetation branch which meets a pattern to form a localized state, and a bare-soil state in cases where it is stable in the same parameter range. 

Within the Turing instability range the uniform-vegetation state is unstable to non-uniform perturbations, so that small perturbations will lead to a pattern with a dominant wavelength $\lambda_d$, that is associated with the fastest growing mode of the system $\sigma_c>0$. In the snaking range the uniform-vegetation state is stable, so that small perturbations never grow regardless of their wavelength. Still, we can calculate $\lambda_d$, for which the growth rate curve $\sigma(\lambda_d)$ has its maxima. Since localized states form a domain that is similar to a periodic state, we may assume it selects a wavelength from the periodic patterns that are stable for this parameter values. As the wavelength of $\sigma (\lambda_d)$ grows faster than the other wavelengths we would expect it is the most dominant one, and may have the largest influence on the snaking wavelength. 

It has recently been shown that the structure of homoclinic snaking and its localized states can change drastically depending on the properties of the uniform-vegetation branch that it is bistable with \cite{Zelnik2016desertification}. In particular, the tails that connect the patterned domain with the uniform-vegetation domain can change from being sinusoidal to exponential, at which point the snaking structure breaks down. It has also been observed that the wavelength of the localized states changes dramatically inside the snaking range, as part of the transition between types of tails. This behavior can be detected from the eigenvalues of the spatial dynamics version of the model, and we use these to define the wavelength of the tail the connects the uniform-vegetation domain to the periodic domain (see methods for more details). If this length scale, which we call the tail wavelength $\lambda_t$, has an affect on the snaking wavelength, we can say that the uniform-vegetation branch is selecting the snaking wavelength.

Finally, if the system is tristable, then both bare-soil and uniform-vegetation are stable in the same range where the localized state exist. In this case there may be a point where a front between these two uniform state will exist and be stationary, termed the Maxwell point. If the Maxwell point is close to the snaking range then each period inside the localized state, which changes between low vegetation and high vegetation, becomes similar to two opposing stationary fronts, connecting between the bare-soil and uniform vegetation states. In this case the length scale of such a front will have an affect on the snaking wavelength, since each period will be approximately made up of two such fronts.

\begin{figure}
    \centering
    \includegraphics[scale=0.4]{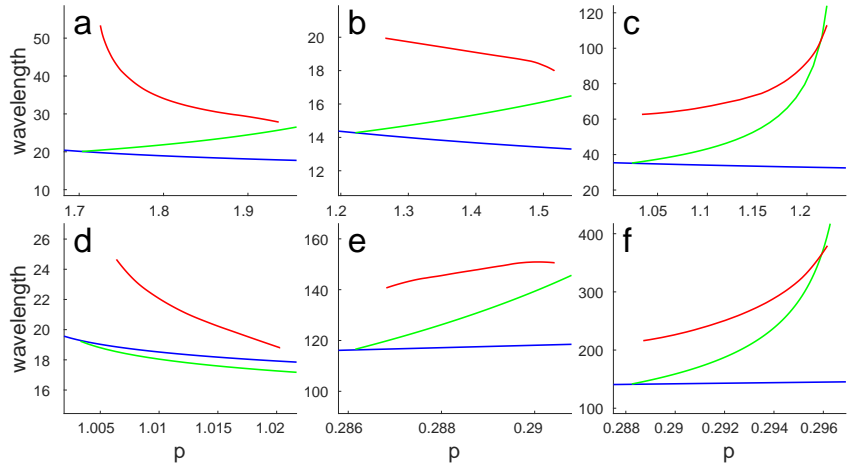}
    \caption{Comparison of three length scales in different values of precipitation $p$: the snaking wavelength $\lambda_s$ in red, the dominant wavelength $\lambda_d$ in blue, and the tail wavelength $\lambda_t$ in green. \textbf{(a,d)} shows large $\lambda_s$ due to proximity to the Maxwell point, while \textbf{(c,f)} show large $\lambda_s$ due to the large $\lambda_t$ (see main text for details).
    The results were made using:
    \textbf{(a,b,c)} G model with parameter sets (3,1,4) respectively, \textbf{(d)} using the K model, and \textbf{(e,f)} with the R model using parameter sets (1,2) respectively.}
    \label{fig:snaking}
\end{figure}

We can now consider how these different effects might choose the snaking wavelength $\lambda_s$. In Fig. \ref{fig:snaking} we plot $\lambda_s$ inside the snaking range, against the dominant wavelength $\lambda_d$ and the sinusoidal tail wavelength $\lambda_t$. As can be seen, $\lambda_s$ generally has values that are similar to $\lambda_d$, and in Fig. \ref{fig:snaking}b,e the trend is also similar. However, in Fig. \ref{fig:snaking}c,f $\lambda_s$ grows for high values of $p$, following a similar trend of the tail wavelength $\lambda_t$. This occurs since when the spatial eigenvalues become real valued close to the snaking range, the wavelength of the sinusoidal tail diverges, and the snaking wavelength follows it. In this case the tail select the wavelength by forcing a large wavelength on the outskirts of the patterned domain, which has an affect on the whole pattern.

A similar situation occurs in Fig. \ref{fig:snaking}a,d where $\lambda_s$ grows for low values of $p$. In these cases however, neither $\lambda_d$ nor $\lambda_t$ show a similar trend, and cannot account for this growth. The change here occurs since the snaking range is close enough to the Maxwell point, so that each period in the localized state is approximately made up of two opposite fronts. Since two such fronts, if they could be realized, could have any distance between them, it follows that each period becomes larger (with a wider gap in the middle) when the localized solution is closer to the Maxwell point. The snaking wavelength therefore becomes larger as it gets closer to the Maxwell point.

We note that while the large-wavelength localized states associated with the first case is typically made of small gaps in a uniform-vegetation background, the second case selects for a large-wavelength pattern that is more similar to solitary peaks. Finally, the second case can only occur if the system exhibits tristability since the Maxwell point occurs inside a bistability range of the uniform states. On the other hand, while we have only seen the first case appear in parameters and models where the uniform-vegetation branch bifurcates supercritically from the bare-soil branch, we know of no reason why other bifurcation structures cannot exhibit this phenomenon as well.

\subsection{Wavelength selection in random conditions}

Real ecosystems are influenced by stochastic environmental conditions and external disturbances such as grazing or fires. This raises the question of what are the typical periodicities we will observe in nature, considering such random perturbations. In such cases, inside the snaking range periodic states are not often selected, but rather different kinds of localized states, all with the same snaking wavelength inside the patterned domain. We therefore choose to define an effective wavelength, which is the system size divided by the number of peaks/gaps in the system. This definition coincides with the pattern wavelength for periodic states, and is more informative for localized states inside the snaking range.

We will consider two types of scenarios where random processes affect the pattern that is selected, convergence after random initial conditions (RIC), and repeated local disturbances (RLD). In the first scenario we start the biomass with a random uniform distribution with no spatial correlation, with biomass values ranging between zero (bare-soil) and the typical maximal biomass for patterns in the given parameters (see Methods). We then integrate the system in time until it converges to a given state.

In the second scenario we begin with random initial conditions as in the previous scenario, and then repeatedly disturb the system by selecting a region of size $s$ in a random location and set the biomass in this region to zero, followed by a given relaxation time $\tau$ for the system to recover.

\begin{figure}
    \centering
    \includegraphics[scale=0.4]{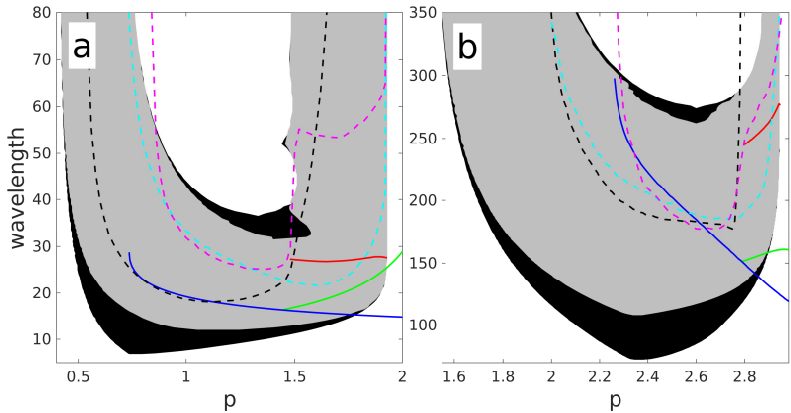}
    \caption{The effective wavelength of solutions in noisy environments as a function of the precipitation $p$ for \textbf{(a)} the G model with parameter set 2 and \textbf{(b)} the R model with parameter set 3. The effective wavelength due to random initial conditions (black) and due to repeated small and large disturbances (cyan and magenta, respectively) are shown in dashed lines. $\lambda_s$,$\lambda_t$, $\lambda_d$ (red,green and blue respectively) are shown in solid curves, similarly to Fig. \ref{fig:snaking}. All the curves are drawn on the background of a Busse balloon where the existence range of periodic solutions is shaded black, and the stability of periodic solutions is shaded grey. The small and large disturbance sizes used were $s=(10,40)$ for the G model and $s=(120,320)$ for the R model, respectively.}
    \label{fig:balloons}
\end{figure}

In Fig. \ref{fig:balloons}, for both the G and R model, we plot the results of the two random scenarios, together with the curves shown in Fig. \ref{fig:snaking}, against the Busse balloon that shows the overall existence and stability of periodic solutions. While there are clear differences between the two models, among them the higher snaking wavelength compared to other length scales in the R model, we can note some clear trends.

First, outside of the snaking range the curve of effective wavelength for RIC (in black) follows $\lambda_d$ quite well, in particular for the G model. Per contra, inside the snaking range the RIC curve shoots up, signifying a few localized gaps in a largely uniform-vegetation domain. We notice that the RIC curve is not limited by the existence range of $\lambda_d$, so that even where uniform-vegetation does not exist we can still get periodic solutions in this way.

Second, we can distinguish between three different regimes of behavior when applying local disturbances. For low $p$, large disturbances (magenta curve) tend to select large wavelength solutions (since not much of the biomass remains) compared to small disturbances (cyan curve). In mid-ranges of $p$, the disturbances size is not very significant, as both curves  largely fall into the same values. Finally, in high $p$ values inside the snaking range, small disturbances select periodic solutions with a small wavelength (compared to the snaking wavelength), while larger disturbances lead to localized solutions with an effective wavelength that is larger then the snaking wavelength.

\begin{figure}
    \centering
    \includegraphics[scale=0.4]{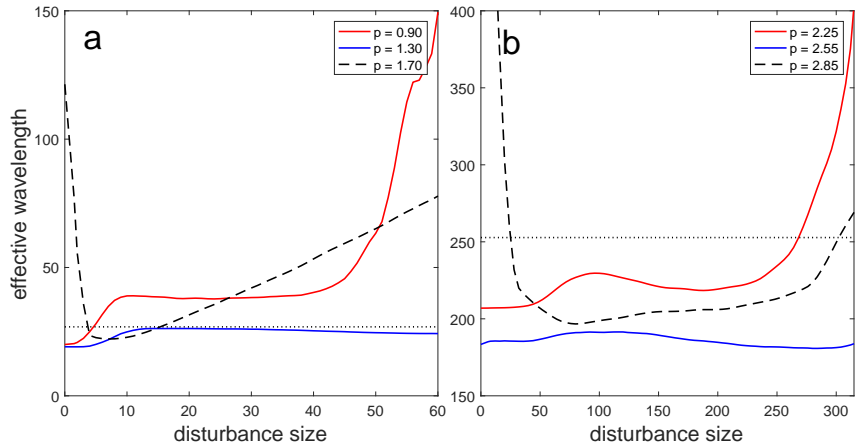}
    \caption{Scaling of the effective wavelength, measured after repeated local disturbances, with respect to the disturbance size, for \textbf{(a)} the G model with parameter set 2 and \textbf{(b)} the R model with parameter set 3. The black dashed curve denotes $p$ value inside the snaking range, while the solid blue and red curves are outside the snaking range. The horizontal dotted line denotes the snaking wavelength for the appropriate value of $p$ ((a) $p=1.70$,(b) $p=2.85$). }
    \label{fig:scaling}
\end{figure}

The behaviour of these three regimes is clarified in Fig. \ref{fig:scaling}, where we look at how the effective wavelength changes as we increase the size of disturbances. Outside the snaking range, small disturbances do not significantly influence the final state of the system. Yet, at an intermediate range of disturbance size the wavelength increases until it settles at a new value. Subsequently it decreases very slowly as $s$ increases. Finally, at low value of $p$, large enough disturbances bring a divergence of the wavelength. In contrast, in the snaking range small disturbances already have a major effect and quickly push the effective wavelength down, selecting periodic solutions with a wavelength that is smaller than $\lambda_s$. At some critical disturbances size (that is smaller than $\lambda_s$ but comparable to it) the trend reverses, and the wavelength becomes larger with increasing disturbance size. We note that this increase is surprisingly constant for the G model, even as the patterns change form periodic to localized state, while for the R model there is a change in slope at high disturbance size.

\section{Discussion}
\label{sec:discussion}
We have shown that the nature of wavelength selection in dryland vegetation ecosystem is complex and affected by various properties of the system. Even though the alternative uniform states are stable in the snaking range, we can identify a relation of the snaking wavelength to the dominant wavelength that is associated with the highest growth mode. We observe that the solutions along the snaking branches select a wavelength within the stable region of periodic solution of the Busse balloon. 

However, the wavelength of the snaking solution changes considerably when there is interference with the Maxwell point of the uniform states. When the Maxwell point is close to the snaking range we see that the snaking wavelength diverges. A similar phenomenon, albeit for a different reason, occurs when the snaking range approaches a a uniform-vegetation state with real spatial eigenvalues. The lack of oscillatory tails prevents the stabilization of localized states, and therefore the wavelength of the snaking solution diverges towards the point in which the spatial eigenvalues become real valued. We thus identify three different length scales that may affect the selected wavelength of the localized state. Whether such a relation can be explicitly shown remains an open question, since deriving an analytical form for localized solutions of such models has yet to be done. It is also unclear how these different effects interact, and in particular if the snaking wavelength can be affected by the Maxwell point and real spatial eigenvalues for the same parameter values.

Applying random local disturbances on the systems under study, we have identified three different responses of the system, corresponding to different precipitation regimes. In the mid-range $p$ values the disturbance regime tends to have a constant effect irregardless of the disturbances size, once over a minimal threshold of disturbance size. This contrasts with both low and high values of $p$ (corresponding approximately to the two bistability ranges) where large disturbances lead to a high effective wavelength in both cases. It is in the high $p$ snaking range however that we see that even very small disturbances can have a significant effect, with a strong inverse relation between effective wavelength and disturbances size. In general, we find similarities in the behaviour of the two models in the scaling of the effective wavelength with respect to the size of the disturbance, at different precipitation value. 

Notwithstanding the challenge of understanding the wavelength selection under stochastic conditions, our results suggests that the underlying characteristic wavelengths of the system can disclose some information on their responses. Further research is needed in order to understand the responses of those systems to disturbances. It would be interesting to compare these results with other models of dryland vegetation, and in general to pattern forming models of other systems. The addition of two-dimensional systems, while bringing higher complexity and numerical challenges, would also be highly beneficial. We thus hope that we open the door for more interesting questions and answers regarding the dynamics which lead to different wavelengths in pattern-forming systems.

\section*{Acknowledgement}
\label{sec:acknowledgement}
We wish to thank Ehud Meron, Hannes Uecker, Golan Bel and Ulrike Feudel for helpful discussions.

\bibliographystyle{unsrt}
\bibliography{WLS}

\end{document}